\PassOptionsToPackage{textwidth=9em, textsize=small}{todonotes}

\documentclass{article} % For LaTeX2e
\usepackage{iclr2020_conference,times}

\usepackage{amsmath,amsfonts,bm}

\def\eqref#1{equation~\ref{#1}}
\def\1{\bm{1}}

\DeclareMathAlphabet{\mathsfit}{\encodingdefault}{\sfdefault}{m}{sl}
\SetMathAlphabet{\mathsfit}{bold}{\encodingdefault}{\sfdefault}{bx}{n}

\usepackage{url}
\usepackage{graphicx}
\usepackage{todonotes}
\usepackage{xspace}
\usepackage{comment}
\usepackage{enumitem}

\usepackage{booktabs}   %% For formal tables:
\usepackage{subcaption} %% For complex figures with subfigures/subcaptions
\usepackage{graphics}
\usepackage{array}
\usepackage{proof}

\usepackage{caption}
\usepackage{hyperref}

\newcommand{\lambdaNet}{\textsc{LambdaNet}}

\title{\lambdaNet{}: Probabilistic Type Inference\\ using Graph Neural Networks}

\author{Jiayi Wei, Maruth Goyal, Greg Durrett, Isil Dillig\\
Department of Computer Science\\
University of Texas at Austin\\
\texttt{\{jiayi, maruth, gdurrett, isil\}@cs.utexas.edu}\\
}

\newcommand{\rev}[2]{#1}
\newcommand{\revSilent}[1]{#1}

\newcommand{\code}[1]{\texttt{#1}}

\newcommand{\heading}[1]{\textbf{#1}\hspace{0.5em}}

\newcommand{\cSharp}{C$^\sharp$}
\newcommand{\concat}{\mathbin{\|}}

\newcommand{\tyvar}{\tau}

\newcommand{\pgraph}{type dependency graph\xspace}

\newcolumntype{L}{l<{\ttfamily}}
\newcolumntype{T}{l<{\ttfamily}<{\hspace{2em}}}
\newcolumntype{R}{r<{\text}}

\newcommand{\weblink}[1]{\href{#1}{\texttt{#1}}}

\iclrfinalcopy % Uncomment for camera-ready version, but NOT for submission.
\begin{document}

\maketitle

\begin{abstract}
As \emph{gradual typing} becomes increasingly popular in languages like Python and TypeScript, there is a growing need to infer type annotations automatically. While type annotations help with tasks like code completion and static error catching, these annotations cannot be fully determined by compilers and are tedious to annotate by hand. This paper proposes a probabilistic type inference scheme for TypeScript based on a graph neural network. Our approach first uses lightweight source code analysis to generate a program abstraction called a \emph{type dependency graph}, which links type variables with logical constraints as well as name and usage information. Given this program abstraction, we then use a graph neural network to propagate information between related type variables and eventually make type predictions. Our neural architecture can predict both standard types, like \code{number} or \code{string}, as well as user-defined types that have not been encountered during training.
Our experimental results show that our approach outperforms prior work in this space by $14\%$ (absolute) on library types, while having the ability to make type predictions that are out of scope for existing techniques. 

\end{abstract}

\section{Introduction}\label{sec:intro}

% Give background and explain why this problem is important.
Dynamically typed  languages like Python, Ruby, and Javascript have  gained enormous popularity over the last decade, yet their lack of a static type system comes with certain disadvantages in terms of  maintainability~\citep{type-maintain}, the ability to catch  errors at compile time, and code completion support~\citep{type-or-not}. \emph{Gradual typing} can address these shortcomings: program variables have \emph{optional} type annotations so that the type system can perform static type checking whenever possible~\citep{Siek2007GradualTF, chung2018kafka}. Support for gradual typing now exists in many popular programming languages~\citep{understanding-typescript, vitousek2014design}, but due to their heavy use of dynamic language  constructs and the absence of principal types~\citep{ancona2004principal}, compilers cannot  perform type inference using standard algorithms from the programming languages community~\citep{understanding-typescript, hindley-milner, pierce2000local}, and manually adding type annotations to existing codebases is a tedious and error-prone task. As a result, legacy programs in these languages do not reap all the benefits of gradual typing.

% Introduce the ML for type inference problem and explain why we focus on TypeScript.
To reduce the human effort involved in transitioning from untyped to statically typed code, this work focuses on a learning-based approach to automatically inferring likely type annotations for untyped (or partially typed) codebases. Specifically, we target TypeScript, a gradually-typed variant of Javascript for which plenty of training data is available in terms of type-annotated programs. While there has been some prior work on inferring type annotations for TypeScript using  machine learning \citep{DeepTyper, JSNice}, prior work in this space has several shortcomings. First, inference is restricted to a finite dictionary of types that have been observed during training time---i.e., they cannot predict any user-defined data types. Second, even without considering user-defined types, the accuracy of these systems is relatively low, with the current state-of-the-art achieving 56.9\% accuracy for primitive/library types~\citep{DeepTyper}. Finally, these techniques can produce inconsistent results in that they may predict different types for different token-level occurrences of the same variable.

% pgraph
In this paper, we propose a new probabilistic type inference algorithm for TypeScript to address these shortcomings using a graph neural network architecture (GNN)~\citep{GAT, Gated-GNN, mou2016convolutional}. Our method uses lightweight source code analysis to transform the program into a new representation called a \emph{\pgraph}, where nodes represent type variables and labeled hyperedges encode relationships between them. In addition to expressing logical constraints (e.g., subtyping relations) as in traditional type inference, a \pgraph  also incorporates contextual hints involving naming and variable usage.

Given such a \pgraph,  our approach uses a GNN to compute a vector embedding for each type variable and then performs type prediction  using a pointer-network-like architecture~\citep{vinyals_pointer_2015}. The graph neural network itself requires handling a variety of hyperedge types---some with variable numbers of arguments---for which we define appropriate graph propagation operators. Our prediction layer compares the vector embedding of a type variable with vector representations of candidate types, allowing us to flexibly handle user-defined types that have not been observed during training. 
%as these types also exist in our graph and  have representations we can consult.
Moreover, our model predicts consistent type assignments by construction because it makes variable-level rather than token-level predictions.
%Specifically, our prediction layer takes as input the vector embedding for a type variable $\alpha$ and a candidate type $\tau$ in the prediction space and computes a compatibility store between $\alpha$ and $\tau$. Because user-defined types also correspond to type variables in our \pgraph, this gives us a natural candidate type embedding and allows us to predict user-defined types in addition to types encountered during training. Moreover, our model predicts consistent results by construction because it predicts at variable-level instead of token-level.

We implemented our new architecture as a tool called \lambdaNet{} and evaluated its performance on real-world TypeScript projects  from Github. When only predicting library types, \lambdaNet{} has a top1 accuracy of 
$75.6\%$, achieving a significant improvement over DeepTyper ($61.5\%$). In terms of overall accuracy (including user-defined types), \lambdaNet{} achieves a top1 accuracy of around $64.2\%$, which is $55.2\%$ (absolute) higher than the TypeScript compiler.

\heading{Contributions.} This paper makes the following contributions: (1)  We propose a probabilistic type inference algorithm for TypeScript that uses deep learning to make predictions from the \pgraph representation of the program. (2) We describe a technique for computing vector embeddings of type variables using GNNs and propose a  pointer-network-like method to predict user-defined types. (3) We experimentally evaluate our approach on hundreds of real-world TypeScript projects and show that our method significantly improves upon prior work.

%\section{Background}

%In this section, we provide a motivating example to illustrate the challenges of type inference for TypeScript and briefly survey prior work in this area.

% Give an example program and briefly mention the kind of reasoning required to solve the prediction task. Briefly describe how our new approach addresses these challenges.
\section{Motivating Example and Problem Setting} 

\begin{figure}
\vspace{-0.2in}
  \centering
  \begin{center}
   \includegraphics[width=0.7\linewidth]{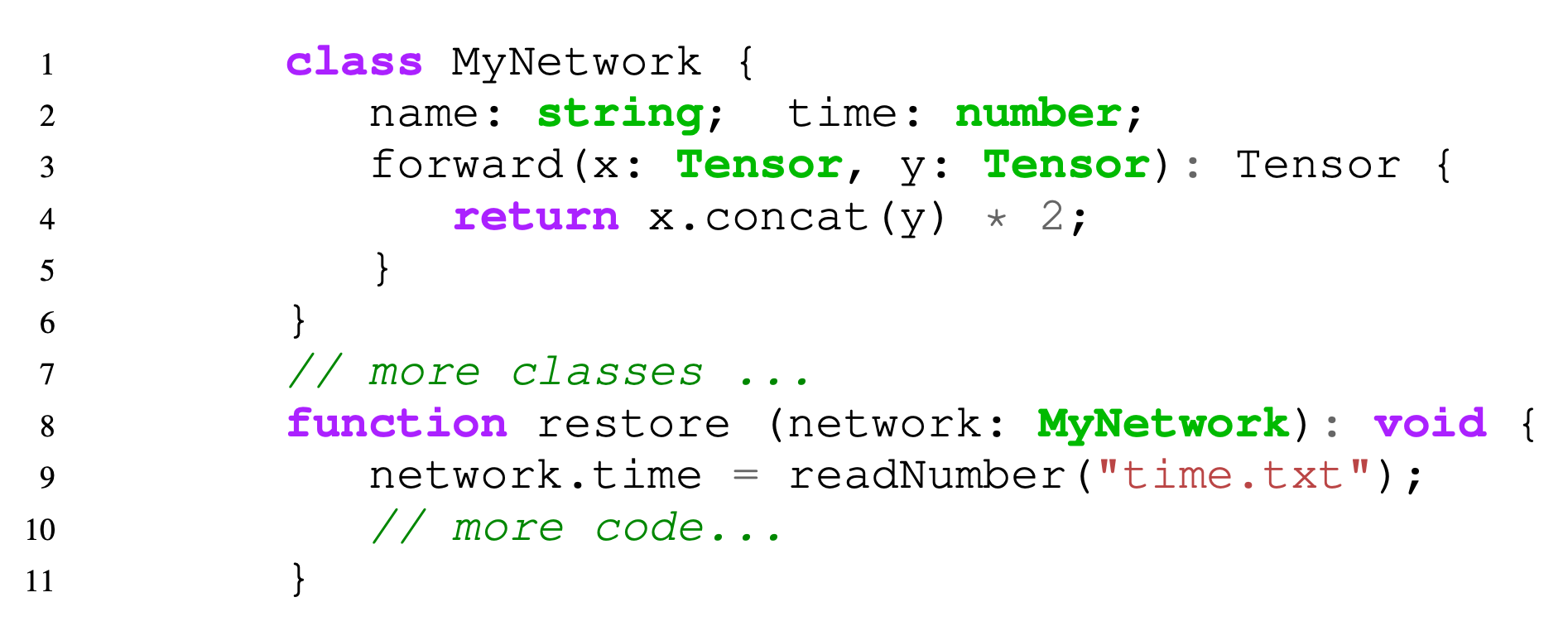}

   % \small
   % \begin{minipage}{0.8\textwidth}
   % \begin{minted}[escapeinside=||, fontsize=\footnotesize, linenos]{TypeScript}
   %    class MyNetwork {
   %       name: string;  time: number;
   %       forward(x: Tensor, y: Tensor): Tensor {
   %          return x.concat(y) * 2; 
   %       }
   %    }  
   %    // more classes ...
   %    function restore (network: MyNetwork): void {
   %       network.time = readNumber("time.txt");
   %       // more code...
   %    }
   % \end{minted}
   % \end{minipage}
%   \vspace{-0.2in}
  
    \caption{\label{fig:intro-example} A motivating example: Given an unannotated version of this TypeScript program, a traditional rule-based type inference algorithm cannot soundly deduce the true type annotations (shown in green).}
  
   \vspace{-0.5em}
   \end{center}
\end{figure}

Figure~\ref{fig:intro-example} shows a (type-annotated) TypeScript program. Our goal in this work is to infer the types shown in the figure, given an \emph{unannotated} version of this code. We now justify various aspects of our solution using this example.

\heading{Typing constraints.} The use of certain functions/operators in  Figure~\ref{fig:intro-example} imposes hard constraints on the types that can be assigned to program variables. For example, in the {\tt forward} function,  variables {\tt x}, {\tt y} must be assigned a type that supports a \code{concat} operation; hence, {\tt x}, {\tt y} could have types like {\tt string}, {\tt array}, or {\tt Tensor}, but not, for example, {\tt boolean}. This observation motivates us to incorporate typing constraints into our model.

\heading{Contextual hints.} Typing constraints are not always sufficient for determining the intended type of a variable. For example, for variable  \code{network}  in function \code{restore}, the typing constraints require \code{network}'s type to be a class with a field called \code{time}, but there can be \emph{many} classes that have such an attribute (e.g., \code{Date}). However, the similarity between the variable name \code{network} and  the class name  \code{MyNetwork} hints that \code{network} might have type \code{MyNetwork}. Based on this belief, we can further propagate the return type  of the library function \code{readNumber} (assuming we know it is \code{number})  to infer that the type of the \code{time} field in \code{MyNetwork}  is likely to be \code{number}.

\heading{Need for \pgraph.} There are many ways to view programs---e.g., as token sequences, abstract syntax trees, control flow graphs, etc. However, none of these representations is particularly helpful for  inferring the most likely type annotations. 
%As illustrated earlier, the main sources of information for type inference include typing constraints and contextual hints.
Thus, our method uses static analysis to infer  a set of \emph{predicates} that are relevant to the type inference problem and represents these predicates using a  program abstraction called the \emph{\pgraph}.

\heading{Handling user-defined types.} As mentioned in Section~\ref{sec:intro}, prior techniques can only predict types seen during training. However, the code from Figure~\ref{fig:intro-example} defines its own class called {\tt MyNetwork} and later uses a variables of type {\tt MyNetwork} in the {\tt restore} method. A successful model for this task therefore must dynamically make inferences about user-defined types based on their definitions.

\subsection{Problem Setting}
\label{sec:formulation}
Our goal is to train a type inference model that can take as input an entirely (or partially) unannotated TypeScript project $g$ and output a probability distribution of types for each missing annotation. The prediction space is $\mathcal{Y}(g) = \mathcal{Y}_{\text{lib}} \cup \mathcal{Y}_{\text{user}}(g)$, where $\mathcal{Y}_{\text{user}}(g)$ is the set of all user-defined types (classes/interfaces) declared within $g$, and $\mathcal{Y}_{\text{lib}}$ is a fixed set of commonly-used library types. % that are commonly used across  TypeScript projects. 

Following prior work in this space~\citep{DeepTyper, JSNice, xu2016python}, we limit the scope of our prediction to  non-polymorphic  and non-function types. That is, we do not distinguish between types such as \code{List<T>}, \code{List<number>}, \code{List<string>} etc., and consider them all to be of type \code{List}. Similarly, we also collapse function types like \code{number}~$\rightarrow$~\code{string} and \code{string}~$\rightarrow$~\code{string} into a single type called \code{Function}. We leave the extension of predicting structured types as future work.

%\heading{Modeling Framework}
% For each TypeScript program $g$, let us denote the set of its missing type annotations (e.g., the types of local variables, function parameters, and class attributes, etc.) as $\mathcal{A}(g)$.
%  Our model takes the form $P(Y|g)$
%  where $Y$ denotes a type assignment mapping from $\mathcal{A}(g)$ to $\mathcal{Y}(g)$.
%  % = \prod_{x \in \mathcal{A}(g)} P(Y(x)|g)$. 
%  In other words, our model predicts a probability distribution over type assignments to all variables  $\mathcal{A}(g)$ in program $g$.

%We describe the particulars of our model in the following sections. Our model is a graph neural network (GNN) over our type dependency graph instantiated over $g$. This graph has several different fundamental edge types, which we describe how to handle. The final representation of each type node from the GNN is then fed into a pointer layer to model the corresponding type assignment $y_i$.

\begin{figure}[t!]
   \centering
   \vspace{-2em}
   \includegraphics[width=0.8\linewidth]{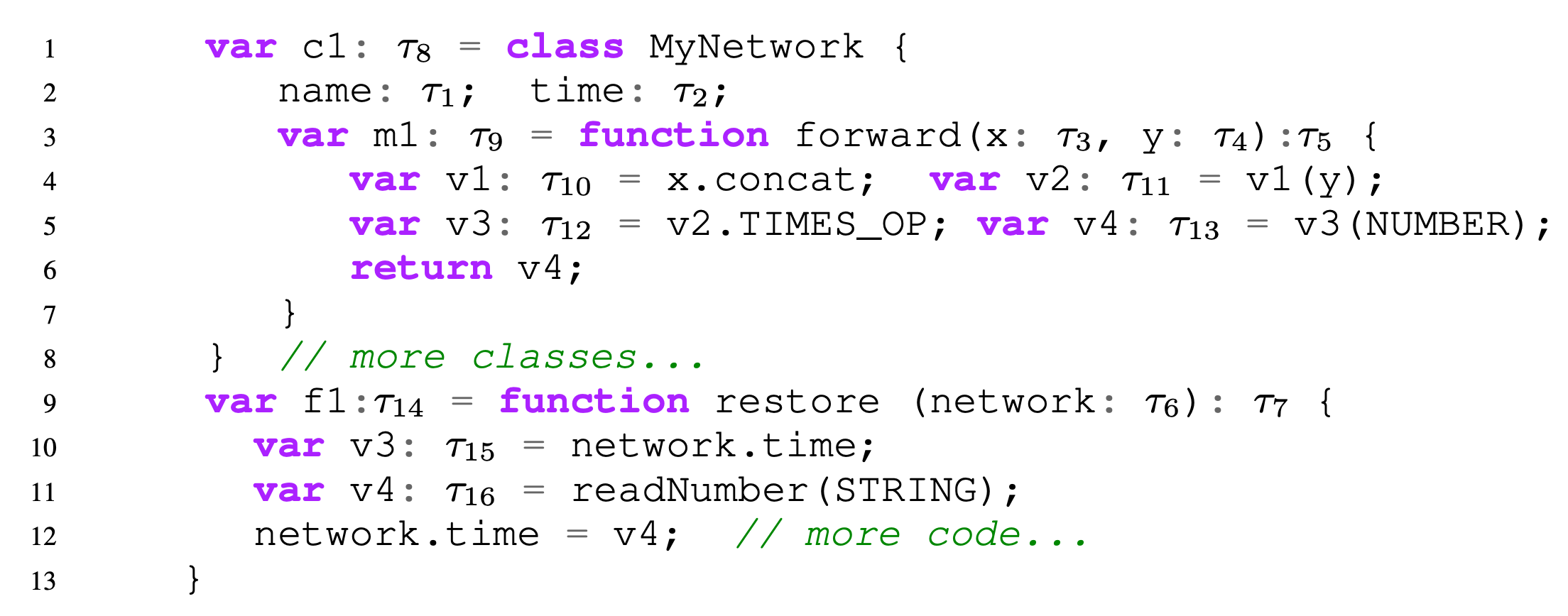}

%    \begin{minipage}{0.7\textwidth}
%    \begin{minted}[escapeinside=||, fontsize=\small{}, linenos]{TypeScript}
%     var c1: |$\tyvar_{8}$| = class MyNetwork {
%        name: |$\tyvar_{1}$|;  time: |$\tyvar_2$|;
%        var m1: |$\tyvar_9$| = function forward(x: |$\tyvar_3$|, y: |$\tyvar_4$|):|$\tyvar_5$| {
%           var v1: |$\tyvar_{10}$| = x.concat;  var v2: |$\tyvar_{11}$| = v1(y);  
%           var v3: |$\tyvar_{12}$| = v2.TIMES_OP; var v4: |$\tyvar_{13}$| = v3(NUMBER); 
%           return v4; 
%        }
%     }  // more classes...
%     var f1:|$\tyvar_{14}$| = function restore (network: |$\tyvar_6$|): |$\tyvar_{7}$| {
%       var v3: |$\tyvar_{15}$| = network.time;  
%       var v4: |$\tyvar_{16}$| = readNumber(STRING);
%       network.time = v4;  // more code...
%    }
%    \end{minted}
%    \vspace{-0.1in}
%   \end{minipage}
  \caption{ An intermediate representation of the (unannotated version) program from Figure~\ref{fig:intro-example}. The $\tau_i$ represent type variables, among which $\tau_8$--$\tau_{16}$ are newly introduced for  intermediate expressions.
  }
  \label{fig:ir-example}
\end{figure}

\begin{figure}
% \vspace{-0.1in}
   \begin{center}
   \includegraphics[width=0.9\linewidth]{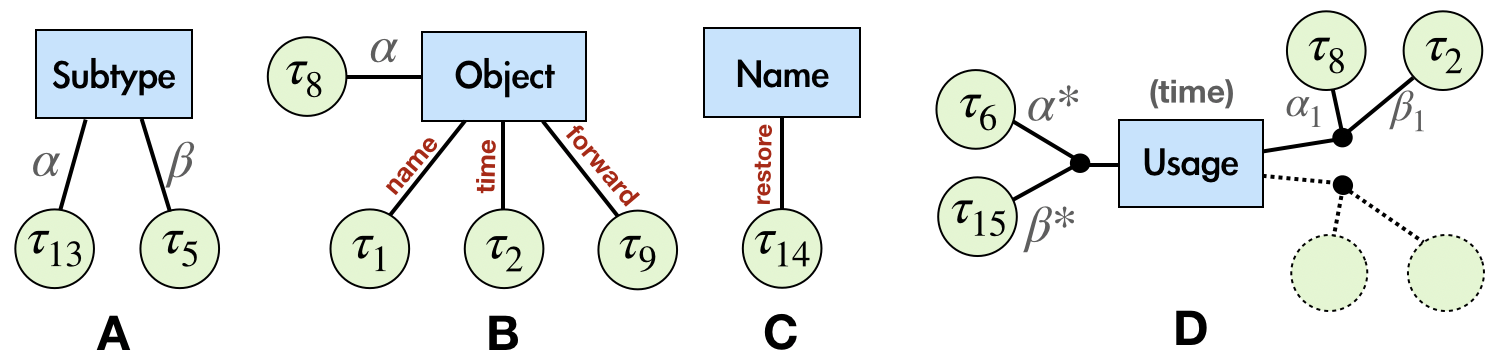}
   \vspace{-0.1in}
   \caption{
      Example hyperedges for Figure~\ref{fig:ir-example}. Edge labels in gray (resp. red) are positional arguments (resp. identifiers).
      \textbf{(A)} The return statement at line 6 induces a subtype relationship between $\tyvar_{13}$ and $\tyvar_5$. 
      \textbf{(B)}  \code{MyNetwork} $\tyvar_8$ declares attributes \code{name} $\tyvar_1$ and \code{time} $\tyvar_2$ and  method \code{forward} $\tyvar_9$.
      \textbf{(C)} $\tyvar_{14}$ is associated with a variable whose named is   \code{restore}.
      \textbf{(D)} \textrm{Usage} hyperedge for line 10 connects $\tyvar_6$ and $\tyvar_{15}$ to  all classes with a \code{time} attribute.
   }
   \label{fig:exampleHyperEdges} 
   \end{center}
  \vspace{-0.8em}
\end{figure}

\section{Type Dependency Graph}

\newcommand{\gSymbol}{\mathcal{G}}
\newcommand{\nLib}{N_{\text{lib}}}
\newcommand{\nProj}{N_\text{proj}}

\newcommand{\setName}[1]{\textsc{#1}}

\begin{table}[!t]
\vspace{-1em}
\small
\caption{Different types of hyperedges used in a \pgraph.}
\label{table:predicates}
   \begin{center}
   \begin{tabular}{LTR}      
\bf{Type} & \bf{Edge} & \bf{Description} \\ \toprule
\multicolumn{3}{c}{\emph{Logical}}\\ \hline 
\textsc{Fixed} & $\text{Bool}(\alpha)$ & $\alpha$ is used as boolean \\
\textsc{Fixed} & $\text{Subtype}(\alpha, \beta)$  & $\alpha$ is a subtype of $\beta$ \\
\textsc{Fixed} & $\text{Assign}(\alpha, \beta)^{\revSilent{\dagger}}$ & $\beta$ is assigned to $\alpha$   \\
   % & $\alpha_1\ \keyword{extends}\ \alpha_2$ & inheritance \\
\textsc{NAry} & $\text{Function}(\alpha, \beta_1, \ldots, \beta_k, \beta^*)$ & $\alpha = (\beta_1, \ldots , \beta_{k}) \rightarrow \beta^*$  \\
\textsc{NAry} & $\text{Call}(\alpha, \beta^*, \beta_1, \ldots, \beta_k)$ & $\alpha = \beta^*(\beta_1, \ldots, \beta_k)$ \\
\textsc{NAry} & $\text{Object}_{l_1,\ldots,l_k}(\alpha, \beta_1, \ldots, \beta_k)$ & $\alpha = \{ l_1: \beta_1, \ldots , l_k: \beta_k \}$ \\ 
\textsc{Fixed} & $\text{Access}_l(\alpha, \beta)$ & $\alpha = \beta.l$ \\
\hline
\multicolumn{3}{c}{\emph{Contextual}}\\ \hline
\textsc{Fixed} & $\text{Name}_{l}(\alpha)$ & $\alpha$ has name $l$ \\
\textsc{Fixed} & $\text{NameSimilar}(\alpha, \beta)$ & $\alpha$, $\beta$ have similar names \\
   
\textsc{NPairs} & $\text{Usage}_l((\alpha^*,\beta^*), (\alpha_1, \beta_1), \ldots, (\alpha_k, \beta_k))$ & usages involving name $l$ \\
\bottomrule\\
  % \multicolumn{3}{r}{($l$ are identifiers \hspace{1em} $\alpha, \beta, \gamma$ are type variables)} \\
   \end{tabular}
   
   \vspace{-0.10in}
   \parbox{0.8\linewidth}{\small $^\dagger$ Although assignment is a special case of a subtype constraint, we differentiate them because these edges appear in different contexts and having uncoupled parameters for these two edge types is beneficial.}
\end{center}

\vspace{-0.15in}
\end{table}

%In this section, we introduce \emph{type dependency graphs} and discuss the semantic and syntactic information they encode.  
A \pgraph $\gSymbol = (N,E)$ is a hypergraph where nodes $N$ represent type variables and labeled hyperedges $E$ encode relationships between them. 
We extract the \pgraph  of a given TypeScript program by performing static analysis on an \emph{intermediate representation} of its source code, which allows us to associate a unique variable  with each program sub-expression.
%and facilitates the extraction of relationships between the types of program expressions. 
As an illustration, Figure~\ref{fig:ir-example} shows the intermediate representation of the code from Figure~\ref{fig:intro-example}.

\begin{comment}
To construct a \pgraph, we need to assign type variables to not only places where a user can put type annotations, but also to all the classes, functions, and intermediate expressions, as they also have their own types. To achieve this, we use light-weight static analysis to transform TypeScript source code into an intermediate representation in which all expressions are explicitly defined through variable declarations. As an example, the program from Figure~\ref{fig:intro-example} is translated into the one shown in Figure~\ref{fig:ir-example}.
\end{comment}

Intuitively, a \pgraph encodes properties of type variables as well as relationships between them. Each hyperedge corresponds to one of the predicates shown in Table~\ref{table:predicates}. We partition our predicates (i.e., hyperedges) into two classes, namely \emph{Logical} and \emph{Contextual}, where the former category can be viewed as imposing hard constraints on type variables and the latter category encodes useful hints extracted from names of variables, functions, and classes. 

% \begin{comment}
% With this intermediate program representation, we then extract different kinds of hyperedges to construct the \pgraph{}. We show a full list of the kinds of edges we use in Table~\ref{table:predicates} and also graphically illustrate 4 examples in Figure~\ref{fig:exampleHyperEdges}.
% \end{comment}

Figure~\ref{fig:exampleHyperEdges} shows some of the hyperedges in the \pgraph $\gSymbol$ extracted from the intermediate representation in Figure~\ref{fig:ir-example}. As shown in Figure~\ref{fig:exampleHyperEdges}(A), our analysis extracts a predicate $\mathrm{Subtype}(\tyvar_{13}, \tyvar_{5})$ from this code because the type variable associated with the returned expression {\tt v4} must be a subtype of the enclosing function's return type. Similarly, as shown in Figure~\ref{fig:exampleHyperEdges}(B), our analysis extracts a predicate $\mathrm{Object}_{\text{name},\text{time}, \text{forward}}(\tyvar_8, \tyvar_1, \tyvar_2, \tyvar_{9})$ because $\tyvar_8$ is an object type whose \code{name}, \code{time}, and \code{forward} members are associated with type variables $\tyvar_1, \tyvar_2, \tyvar_{9}$, respectively.

In contrast to the \rm{Subtype} and \rm{Object} predicates that impose hard constraints on type variables, the next two hyperedges shown in Figure~\ref{fig:exampleHyperEdges} encode contextual clues obtained from variable names. Figure~\ref{fig:exampleHyperEdges}(C) indicates that type variable $\tyvar_{14}$ is associated with an expression named {\tt restore}. While this kind of naming information is invisible to TypeScript's structural type system~\citep{understanding-tt}, it  serves as a useful input feature for our GNN architecture  described in Section~\ref{sec:architecture}.

In addition to storing the unique variable name associated with each type variable, the \pgraph also encodes similarity  between variable and class names. The names of many program variables mimic their types: for example, instances of a class called {\tt MyNetwork} might often be called {\tt network} or {\tt network1}. To capture this correspondence, our \pgraph also contains a hyperedge called \textrm{NameSimilar} that connects type variables $\alpha$ and $\beta$ if their corresponding tokenized names have a non-empty intersection.\footnote{During tokenization, we split identifier names into  tokens based on underscores and  camel case naming. More complex schemes are possible, but we found this simple method to be effective.}

As shown in Table~\ref{table:predicates}, there is a final  type of hyperedge called \textrm{Usage} that facilitates type inference of object types. In particular, if there is an object access \code{var y = x.l}, we extract the predicate $\text{Usage}_l((\tyvar_x,\tyvar_y), (\alpha_1, \beta_1), \ldots, (\alpha_k, \beta_k))$ to connect \code{x} and \code{y}'s type variables with \emph{all} classes $\alpha_i$ that contain an attribute/method $\beta_i$ whose name is \code{l}. Figure~\ref{fig:exampleHyperEdges} shows a \textrm{Usage} hyperedge extracted from the code in Figure~\ref{fig:ir-example}. As we will see in the next section, our GNN architecture utilizes a special attention mechanism to pass information along these usage edges.

\section{Neural Architecture}
\label{sec:architecture}

Our neural architecture for making type predictions consists of two main parts. First, a graph neural network passes information along the type dependency graph to produce a vector-valued embedding for each type variable based on its neighbors. % which include naming information, usage information, and more, as described in the previous section.
Second, a pointer network compares each variable's type embedding to the embedding vectors of candidate types (both  computed from the previous phase) to place a distribution over possible type assignments.

%\subsection{Graph Neural Network Architecture}

% High-level pipeline
Given a \pgraph $\mathcal{G} = (N,E)$, we first to compute a vector embedding $\mathbf{v}_n$ for each $n \in N$ such that these vectors implicitly encode type information. Because our program abstraction  is a graph, a natural choice is to use a graph neural network architecture.  From a high level, this architecture  takes in initial vectors $\mathbf{v}_n^0$ for each node $n$, performs $K$ rounds of message-passing in the graph neural network, and returns the final representation for each type variable.

In more detail, let $\mathbf{v}_n^t$ denote the vector representation of node $n$ at the $t$th  step, where each round  consists of a \emph{message passing} and an \emph{aggregation} step. The message passing step computes a vector-valued update to send to the $j$th argument of each hyper-edge $e \in E$ connecting nodes $p_1, \ldots, p_a$.   Then, once all the messages have been computed, the \emph{aggregation} step computes a new embedding $\mathbf{v}_n^t$ for each $n$ by combining all messages sent to $n$:
\begin{equation*}
\mathbf{m}_{e,p_j}^t = \mathrm{Msg}_{e,j}(\mathbf{v}_{p_1}^{t-1},\ldots,\mathbf{v}_{p_a}^{t-1})\ \ \ \ \ \mathbf{v}_n^t = \mathrm{Aggr}(\mathbf{v}_n^{t-1}, \{\mathbf{m}_{e,n}^t | e \in \mathcal{N}(n)\})
\end{equation*}
Here, $\mathcal{N}$ is the neighborhood function, and $\mathrm{Msg}_e$ denotes a particular neural operation that depends on the type of the edge (\textsc{Fixed}, \textsc{NAry}, or \textsc{NPairs}), which we will describe later. 
%Note that weights are shared between all edges of the same type.
%First, we use an initialization operation to assign each $n \in N$ an initial vector $\vec{v}_n^0$. We can write this as a batched operation $\tens{v}^0 = \tens{Init}_N$, where the $j$th row of $\tens{v}^0$ is $\vec{v}_{n_j}^0$. Then, we perform $K$ times of graph propagation $\tens{v}^t = \tens{Prop}_E(\tens{v}^{t-1})$ for $t=1\ldots K$ and take $\tens{v}^K$ as the final embedding for all type variables. $K$ is a hyperparameter of our architecture. Specifically, the propagation operation is the composition of \emph{message passing} and \emph{aggregation}. For each hyperedge $e \in E$ with arguments $p_1, \ldots, p_a$, \emph{message passing} computes a message the $j$th argument $p_j$, denoted as $\vec{m}_{e,p_j}^t$, using each argument's current embedding. In other words, $\vec{m}_{e,p_j}^t = \tens{Msg}^t_{e,j}(\vec{v}_{p_1}^{t-1}, \ldots, \vec{v}_{p_a}^{t-1})$, and $\emph{Messaging}^t_{e,j}$ is a neural network whose architecture depends on the type of edge $e$ and the position $j$. Once all the messages have been computed, \emph{aggregation} computes a new embedding $v_n^t$ for each $n$ by combining all messages that are sent to $n$, i.e., we have $v_n^t = \tens{Aggr}(\{\vec{m}_{e,n}^t | e \in \text{Neighbour}(n)\})$. We now describe each of these operations in detail.

% Initialization
\heading{Initialization.} In our GNN, nodes correspond to type variables and each type variable is associated either with a program variable or a constant. We refer to nodes representing constants (resp. variables) as \emph{constant (resp. variable) nodes}, and our initialization procedure works differently depending on whether or not $n$ is a constant node. Since the types of each constant are known, we set the initial embedding for each constant node of type $\tau$ (e.g., \texttt{string}) to be a trainable vector $\mathbf{c}_{\tau}$ and do not update it during GNN iterations (i.e., $\forall t, \mathbf{v}_n^t = \mathbf{c}_\tau$). On the other hand, if $n$ is a variable node, then we have no information about its type during initialization; hence, we initialize all variable nodes using a generic trainable initial vector (i.e., they are initialized to the \emph{same} vector but updated to different values during GNN iterations).

% Propagation
\heading{Message passing.} Our $\mathrm{Msg}$ operator depends on the category of edge it corresponds to (see Table~\ref{table:predicates}); however, weights are shared between all instances of the same hyperedge type. In what follows, we describe the neural layer that is used to compute messages for  each  type of hyperedge:

% \todo{GREG: I assume weights are shared between things like "Naming" but not all SIMPLE edges, right? Jiayi: Yes}

\begin{itemize}[leftmargin=*]
   \item \textsc{Fixed}: Since these edges correspond to fixed arity predicates  (and the position of each argument matters), we compute the message of the $j$th argument by  first concatenating the embedding vector of all arguments and then feed the result vector to a 2-layer MLP for the $j$th argument.  In addition, since hyperedges of type \emph{Access}  have an identifier,  we also embed the identifier as a vector and treat it as an extra argument. (We describe the details of identifier embedding later in this section.)
   \item \textsc{NAry}: Since \textsc{NAry}  edges connect a variable number of nodes, we need an architecture that can deal with this challenge. In our current implementation of \lambdaNet{}, we use a simple architecture that is amenable to batching. Specifically, 
   given an \textsc{NAry} edge $\emph{E}_{l_1,\ldots,l_k}(\alpha,\beta_1,\ldots,\beta_k)$ (for \emph{Function} and \emph{Call}, the labels $l_j$ are argument positions), the set of messages for $\alpha$ is computed as $\{\mathrm{MLP}_\alpha(\mathbf{v}_{l_j} \concat \mathbf{v}_{\beta_j})\ |\  j= 1\ldots k \}$, and the message for each $\beta_j$ is computed as $\mathrm{MLP}_\beta(\mathbf{v}_{l_j} \concat \mathbf{v}_{\alpha})$. Observe that we compute $k$ different messages for $\alpha$, and the message for each $\beta_j$ only depends on the vector embedding of $\alpha$ and its position $j$, but not the vector embeddings of other $\beta_j$'s.\footnote{In our current implementation, this is reducible to multiple \textsc{Fixed} edges. However, \textsc{NAry} edges could generally use more complex pooling over their arguments to send more sophisticated messages.}

   \item \textsc{NPairs}: This is a special category  associated with $\text{Usage}_l((\alpha^*,\beta^*), (\alpha_1, \beta_1), \ldots, (\alpha_k, \beta_k))$. Recall that this kind of edge arises from expressions of the form $b=a.l$ and is used to connect $a$ and $b$'s type variables with all classes $\alpha_i$ that contain an attribute/method $\beta_i$ with label $l$. Intuitively, if $a$'s type embedding is very similar to a type $C$, then $b$'s type will likely be the same as $C.l$'s type. Following this reasoning, we use dot-product based attention to compute the messages for $\alpha^*$ and $\beta^*$. Specifically, we use $\alpha^*$ and $\alpha_j$'s as attention keys and $\beta_j$'s as attention values to compute the message for $\beta^*$ (and switch the key-value roles to compute the message for $\alpha^*$):
   \begin{align*}
   \vspace{-0.1in}
   \small
       \mathbf{m}^t_{e,\beta^*} = \sum_j{w_j \mathbf{v}^{t-1}_{\beta_j}} \hspace{2em} \mathbf{w}=\mathrm{softmax}(\mathbf{a}) \hspace{2em} a_j = \mathbf{v}_{\alpha_j} \cdot \mathbf{v}_{\alpha^*} 
   \end{align*}
  % Similarly, we also switch the key-value roles and compute a message for $\alpha^*$.
         \vspace{-0.2in}
\end{itemize}

% Aggregation
\heading{Aggregation.} Recall that the aggregation step combines all messages sent to node $n$ to compute the new embedding $v_n^t$. To achieve this goal,  we use a variant of the attention-based aggregation operator proposed in graph attention networks~\citep{GAT}. 
\begin{equation}\label{eq:aggregation}
\small
    v_n^t = \mathrm{Aggr}(v_n^{t-1}, \{m_{e,n}^t | e \in \mathcal{N}(n)\}) 
    = v_n^{t-1} + \sum_{e \in \mathcal{N}(n)}{w_e \mathbf{M}_1 m_{e,n}^t }
\end{equation}
where $w_e$ is the attention weight for the message coming from edge $e$. Specifically,  the weights $w_e$ are computed as $\mathrm{softmax}(\mathbf{a})$, where $a_e = \mathrm{LeakyReLu}(v_n^{t-1} \cdot \mathbf{M}_2 m_{e,n}^t)$ , and $\mathbf{M}_1$ and $\mathbf{M}_2$ are trainable matrices. Similar to the original GAT architecture, we set the slope of the LeakyReLu to be $0.2$, but we use dot-product to compute the attention weights instead of a linear model.

\heading{Identifier embedding.}
Like in \cite{allamanis2017learning}, we break variable names into word tokens according to camel case and underscore rules and assign a trainable vector for all word tokens that appear more than once in the training set. For all other tokens, unlike \cite{allamanis2017learning}, which maps them all into one single \code{<Unknown>} token, we randomly mapped them into one of the \code{<Unknown-i>} tokens, where $i$ ranges from 0 to 50 in our current implementation. This mapping is randomly constructed every time we run the GNN and hence helps our neural networks to distinguish different tokens even if they are rare tokens. We train these identifier embeddings end-to-end along with the rest of our architecture.

\heading{Prediction Layer.}
For each type variable $n$ and each candidate type $c \in \mathcal{Y}(g)$, we use a MLP to compute a compatibility score $s_{n,c} = \mathrm{MLP}(\mathbf{v}_n,\mathbf{u}_c)$, where $\mathbf{u}_c$ is the embedding vector for $c$. If $c \in \mathcal{Y}_{\text{lib}}$, $\mathbf{v}_c$ is a trainable vector for each library type $c$; if $c \in \mathcal{Y}_{\text{user}}(g)$, then it corresponds to a node $n_c$ in the \pgraph of $g$, so we just use the embedding vector for $n_c$ and set $\mathbf{u}_c = \mathbf{v}_{n_c}$. Formally, this approach looks like a pointer network \citep{vinyals_pointer_2015}, where we use the embeddings computed during the forward pass to predict ``pointers'' to those types.

Given these compatibility scores, we apply a softmax layer to turn them into a probability distribution. i.e., $P_n(c|g) = \exp(s_{n,c}) / \sum_{c'}{\exp(s_{n,c'})}$. During test time, we max over the probabilities to compute the most likely (or top-N) type assignments.

% \begin{comment}
% \subsection{Implementation}

% We implemented \lambdaNet{} in Scala, using the popular Java high performance Tensor library \emph{Nd4j}~\citep{?}. This gives us more flexibility when designing the message passing layers for dealing with hyperedges. We perform graph level batching when computing the messages.

% For parsing TypeScript source code, we use the TypeScript compiler API to pre-process TypeScript programs into Scala ASTs.

% \todo{What else do we need to put here?}

% \heading{Hyperparameters} todo here...

% \subsection{Training}

% \todo{I'm putting this here, not sure where it should go but it's weird to say training before the model} We train our model with standard supervised learning, maximizing the log likelihood of a labeled set of programs $G_{\text{data}} = [(g_i,Y_i^*)]_{i=1}^n$. \todo{are these fully labeled or partially labeled? should specify that here} During learning, we maximize the likelihood of the gold annotations:
% \newcommand{\expectation}{\mathop{\mathbb{E}}}
% \begin{equation}
%     \sum_{(g, Y^*) \in G_{\text{data}}}\left[
%         \sum_{y_j^* \in Y^*}{
%             \log(P(y_j^*|g))
%         }
%     \right]
% \end{equation}
% \end{comment}

\section{Evaluation}

In this section, we describe the results of our experimental evaluation, which is designed to answer the following questions:  (1) How does our approach compare to previous work? (2) How well can our model predict user-defined types? (3) How useful is each of our model's components?

\heading{Dataset.} Similar to \cite{DeepTyper}, we train and evaluate our model on popular open-source TypeScript projects taken from Github. Specifically, we collect 300 popular TypeScript projects from Github that contain between $500$ to $10,000$ lines of code and where at least $10\%$ of type annotations are user-defined types. \rev{Note that each project typically contains hundreds to thousands of type variables to predict, and these projects in total contain about 1.2 million lines of TypeScript code.}{Added total amount of TypeScript code} Among these $300$ projects, we use $60$ for testing, $40$ for validation, and the remainder for training. 

\rev{\heading{Code Duplication.} We ran jscpd\footnote{A popular code duplication detection tool, available at \weblink{https://github.com/kucherenko/jscpd}.} on our entire data set and found that only 2.7\% of the code is duplicated. Furthermore, most of these duplicates are intra-project. Thus, we believe that code duplication is not a severe problem in our dataset.}{Added information about code duplication}

 %Among the 998 TypeScript projects,\todo{is this the same set they used? unclear} we filtered out projects that are too small ($\textless$500 LOC) or too big ($\textgreater$10K LOC);\todo{why this filtering? would be best if we can argue that these projects are weird and not that our method doesn't scale} this left us with about 500 projects. We then randomly selected 60 of them as our testing set. Among the remaining projects, we also filtered out projects whose annotation set contains too few user-defined types ($\textless$10\%); this left us with a training set of 200 projects and a validation set of 40 projects. \greg{Note that each project contains hundreds or thousands of type variables to predict}

\heading{Preprocessing.} Because some of the projects in our benchmark suite are only sparsely type annotated, we augment our labeled training data by using the forward type inference functionality provided by the TypeScript compiler\rev{.}{Explains why we need forward inference in our training set}\footnote{Like in many modern programming languages with forward type inference (e.g., Scala, C\#, Swift), a TypeScript programmer does not need to annotate every definition in order to fully specify the types of a program. Instead, they only need to annotate some “key places” (e.g., function parameters and return types, class members) and let the forward inference algorithm to figure out the rest of the types. Therefore, in our training set, we can keep the user annotations on these key places and run the TS compiler to recover these implicitly specified types as additional labels.} 
The compiler cannot infer the type of every variable and leaves many labeled as \code{any} during failed inference; thus, we exclude \code{any} labels in our data set.  Furthermore, at test time, we  evaluate our technique only on annotations that are manually added by developers. This is the same methodology used by \cite{DeepTyper}, and, since developers  often add annotations where code is most unclear,  this  constitutes a challenging setting for type prediction.

%Since the TypeScript compiler supports a restricted form of forward type inference, to improve training, we not only train our model on type annotations that developers have manually added, but also use the compiler to infer additional type annotations for all intermediate expressions. We exclude all instances of \code{any} in our annotation set because they are often used by the compiler to signal failing to infer an accurate type. As in \cite{DeepTyper}, we only evaluate on annotations that are manually added by the developers at test time. \greg{Note that developers will often add annotations where code is most unclear, so this constitutes a challenging setting.}

\heading{Prediction Space.} As mentioned in Section~\ref{sec:formulation}, \revSilent{our approach takes an entire TypeScript project $g$ as its input}, and the corresponding type prediction space is $\mathcal{Y}(g) = \mathcal{Y}_{\text{lib}} \cup \mathcal{Y}_{\text{user}}(g)$. In our experiments, we set $\mathcal{Y}_\text{user}(g)$ to be all classes/interfaces defined in $g$ (except when comparing with DeepTyper, where we set $\mathcal{Y}_{\text{user}}(g)$ to be empty), and for $\mathcal{Y}_\text{lib}$, we select the top-100 most common types in our training set. Note that this covers $98\%$ (resp. $97.5\%$) of the non-\code{any} annotations for the training (resp. test) set.

\rev{\heading{Hyperparameters} We selected hyperparameters by tuning on a validation set as we were developing our model.}{Made hyperparameters a spearate minisection} We use 32-dimensional type embedding vectors, and all MLP transformations in our model use one hidden layer of 32 units, except the MLP for computing scores in the prediction layer, which uses three hidden layers of sizes 32,16, and 8 (and size 1 for output). GNN message-passing layers from different time steps have independent weights.

We train our model using Adam~\citep{Adam} with default parameters ($\alpha = 0.9$, $\beta = 0.999$) and set the learning rate to be $10^{-3}$ initially but linearly decrease it to $10^{-4}$ until the 30th epoch. We use a weight decay of $10^{-4}$ for regularization and stop the training once the loss on validation set starts to increase (which usually happens around 30 epochs). We use the type annotations from a single project as a minibatch and limit the maximal batch size (via downsampling) to be the median of our training set to prevent any single project from having too much influence.

\heading{Implementation Details.}
We implemented \lambdaNet{} in Scala, building on top of the Java high-performance Tensor library \emph{Nd4j}\citep{nd4j}, and used a custom automatic differentiation library to implement our GNN. Our GNN implementation does not use an adjacency matrix to represent GNN layers; instead, we build the hyperedge connections directly from our \pgraph{} and perform batching when computing the messages for all hyperedges of the same type. 

\revSilent{\heading{Code Repository.} We have made our code publicly available on Github.\footnote{See \weblink{https://github.com/MrVPlusOne/LambdaNet}.}}

\subsection{Comparison with DeepTyper}

% In order to conduct a  meaningful comparison between DeepTyper and {\sc LambdaNet}, we  make two modifications to the model described by \citep{DeepTyper}.  First, the original DeepTyper implementation makes predictions for each occurrence of a variable rather than for each declaration. Thus, if a program variable appears multiple times in the source file, DeepTyper   makes multiple predictions for it. Since type annotations would be added for each variable declaration (rather than for each occurrence), we implemented a variant of DeepTyper that makes a single prediction  for each variable. We refer to this variant as DeepTyper$^\star$. Second, while DeepTyper uses variable names  in its model, it does not tokenize them (e.g., \code{"MyNetwork"} would be treated as a single token rather than \code{["my", "network"]}. However, since {\sc LambdaNet} does tokenize variable names, we also implement a variant of DeepTyper called DeepTyper-Tokenized that treats variable names in the same way as {\sc LambdaNet}.\todo{GREG: probably we'll discuss tokenization of naming in section 5 or earlier, so update this then}

In this experiment, we compare \lambdaNet{}'s performance with DeepTyper~\citep{DeepTyper}, which treats programs as sequences of tokens and uses a bidirectional RNN to make type predictions. Since DeepTyper can only predict types from a fixed vocabulary, we fix both \lambdaNet{} and DeepTyper's prediction space to  $\mathcal{Y}_\text{lib}$ and measure their corresponding top-1 accuracy.

The original DeepTyper model makes predictions for each variable occurrence rather than declaration. In order to conduct a meaningful comparison between DeepTyper and \lambdaNet{}, we implemented a variant of DeepTyper that makes a single prediction  for each variable (by averaging over the RNN internal states of all occurrences of the same variable before making the prediction). \rev{Moreover, for a fair comparison, we made sure both DeepTyper and \lambdaNet{} are using the same improved naming feature that splits words into tokens.}{Clarifies the naming feature in use}

Our main results are summarized below, where the   Declaration (resp. Occurrence) column shows  accuracy per variable declaration (resp. token-level occurrence). 
Note that we obtain occurrence-level accuracy from declaration-level accuracy by weighting each variable by its number of occurrences.

\begin{centering}
\begin{tabular}{lcc}
    \toprule
    \bf{Model} & \multicolumn{2}{c}{\bf{Top1 Accuracy (\%)}} \\
    & \emph{Declaration} & \emph{Occurrence} \\
    \midrule
    DeepTyper & 61.5 & 67.4 \\
    \lambdaNet{}$_\text{lib}$ (K=6) & 75.6 & 77.0  \\
    \bottomrule
\end{tabular}\par
\end{centering}

As we can see from the table, \lambdaNet{} achieves significantly better results compared to DeepTyper. In particular, \lambdaNet{} outperforms DeepTyper by $14.1\%$ (absolute) for declaration-level accuracy and by $9.6\%$ for occurrence-level accuracy.

Note that the accuracy we report for DeepTyper ($67.4\%$) is not directly comparable to the original accuracy reported in~\cite{DeepTyper} ($56.9\%$) for the following reasons. While we perform static analysis and have a strict distinction of library vs.~user-defined types and only evaluate both tools on library type annotations in this experiment, their implementation treat types as tokens and does not have this distinctions. Hence, their model also considers a much larger prediction space consisting of many user-defined types---most of which are never used outside of the project in which they are defined---and is also evaluated on a different set of annotations than ours.

%and its variants, while the last row reports results for {\sc LambdaNet}. For DeepTyper, the second column labeled Declaration shows the results for DeepTyper$^\star$, and, for {\sc LambdaNet}, the column labeled Occurrence shows accuracy as the weighted average of per-variable predictions weighted by the number of occurrences. Thus, Table~\ref{table:comparison} shows results for a total of four different variants of DeepTyper.

%We show the results in Table~\ref{table:comparison}. To separate the effects of architectural improvements from simpler feature improvements, we also compare with two variants of the original DeepTyper implementation. DeepTyper-Token\todo{this is incredibly confusing...are we conflating Token vs. Node here?} uses our tokenization based naming features instead of the naming feature originally described in their paper. While DeepTyper uses a consistency layer to average over the embeddings of all occurrences of the same \emph{identifier}, DeepTyper-Node utilizes our static analysis to average all occurrences of the same \emph{type variable} instead. 

\begin{table}[t]
  \vspace{-1em}
  \centering
  \caption{Accuracy when predicting all types.}
  \label{table:predict-UDT}
  \begin{tabular}{lccc|ccc}
    \toprule
      \bf{Model} & \multicolumn{3}{c}{\bf{Top1 Accuracy (\%)}} & \multicolumn{3}{c}{\bf{Top5 Accuracy (\%)}} \\
      & $\mathcal{Y}_\text{user}$ &  $\mathcal{Y}_\text{lib}$ & \emph{Overall} & $\mathcal{Y}_\text{user}$ &  $\mathcal{Y}_\text{lib}$ & \emph{Overall} \\
      \midrule
      \textsc{TS Compiler} & 2.66 & 14.39 & 8.98 & - & - & - \\
      \textsc{SimilarName} & 24.1 & 0.78 & 15.7 & 42.5 & 3.19 & 28.4 \\
      \lambdaNet{} (K=6) & 53.4 & 66.9 & 64.2 & 77.7 & 86.2 & 84.5 \\
      \bottomrule
  \end{tabular}
%   \vspace{-1em}
  
\end{table}

\subsection{Predicting User-Defined Types}

As mentioned earlier, our approach differs from prior work in that it is capable of predicting user-defined types; thus, in our second experiment,  we extend \lambdaNet{}'s prediction space to also include user-defined types.  However, since such types are not in the prediction space of prior work~\citep{DeepTyper}, we implemented two simpler baselines that can be used to calibrate our model's performance. Our first baseline is the type inference performed by the TypeScript compiler, which is sound but incomplete (i.e., if it infers a type, it is guaranteed to be correct, but it infers type \code{any} for most variables).\footnote{For inferring types from the TypeScript compiler, we use the code provided by \cite{DeepTyper}. We found this method had a slightly lower accuracy than reported in their work.} Our second baseline, called {\sc SimilarName}, is inspired by the similarity between variable names and their corresponding types; it predicts the type of each variable $v$ to be the type whose name shares the most number of common word tokens with $v$. 

The results of this experiment are shown in Table~\ref{table:predict-UDT}, which shows  the top-1 and top-5 accuracy for both user-defined  and library types individually as well as overall accuracy. In terms of overall prediction accuracy, \lambdaNet{} achieves $64.2\%$ for top-1  and $84.5\%$ for top-5, significantly outperforming both baselines. Our results suggest that our fusion of logical and contextual information to predict types is far more effective than rule-based incorporation of these in isolation.

%Since our model now needs to consider a larger set of possible type assignments, we also include evaluation on library types, now in this more challenging setting.

%To calibrate our model's performance, we also implemented two simpler baselines: \textsc{Assignment} uses a simple heuristic based on local assignments. If a variable is locally assigned with expressions whose types are known (e.g., constants or constructor calls), then \textsc{Assignment} simply predicts the corresponding type (if there are more than one types, choosing the most frequently assigned one). \textsc{SimilarName} predicts the type of a type variable $v$ by finding the candidate type whose name shares the most number of common word tokens with $v$. \todo{what about most frequent class? is that one just too bad?} We show the results in Table~\ref{table:predict-UDT}. \greg{Again, say conclusions once we've got them.}

% \begin{table}
%   \caption{Predicting All Types.  }
%   \label{table:predict-UDT}
% %   \begin{center}

%   \begin{tabular}{lccc|ccc}
%       \bf{Model} & \multicolumn{3}{c}{\bf{Top1 Accuracy (\%)}} & \multicolumn{3}{c}{\bf{Top5 Accuracy (\%)}} \\
%       & $\mathcal{Y}_\text{user}$ &  $\mathcal{Y}_\text{lib}$ & \emph{Overall} & $\mathcal{Y}_\text{user}$ &  $\mathcal{Y}_\text{lib}$ & \emph{Overall} \\
%       \hline \\
%       \textsc{TS Compiler} & \xx & \xx & \xx & \xx & \xx & \xx \\
%       \textsc{SimilarName} & 24.1 & 0.78 & 15.7 & 42.5 & 3.19 & 28.4 \\
%       \lambdaNet{} & \xx & \xx & \xx & \xx & \xx & \xx \\
%   \end{tabular}
   
% %   \end{center}
% \end{table}

\subsection{Ablation Study}

\begin{table}
   \begin{center}
%   \vspace{-1em}
   \caption{Performance of different GNN iterations (left) and ablations (right). }
   \label{table:ablation}

   \begin{tabular}{lccc}
   \toprule
      \bf{K} & \multicolumn{3}{c}{\bf{Top1 Accuracy (\%)}} \\
       & $\mathcal{Y}_\text{user}$ & $\mathcal{Y}_\text{lib}$ & \emph{Overall} \\
      \midrule
      6 & 53.4 & 66.9 & 64.2 \\
      4 & 48.4 & 65.5 & 62.0 \\
      2 & 47.3 & 61.7 & 58.8 \\
      1 & 16.8 & 48.2 & 41.9 \\
      0 &  0.0 & 17.0 & 13.6 \\
      \bottomrule
   \end{tabular}
   \quad{}
   \begin{tabular}{lccc}
   \toprule
      \bf{Ablation} & \multicolumn{3}{c}{\bf{Top1 Accuracy (\%)}} \\
       (K = 4) & $\mathcal{Y}_\text{user}$ & $\mathcal{Y}_\text{lib}$ & \emph{Overall} \\
      \midrule
      \lambdaNet{} & 48.4 & 65.5 & 62.0 \\
      No Attention in \textsc{NPair} & 44.1 & 57.6 & 54.9 \\
      No \emph{Contextual} & 27.2 & 52.6 & 47.5 \\
      No \emph{Logical}$^*$ & 24.7 & 39.2 & 36.2 \\
      Simple Aggregation & 40.2 & 66.9 & 61.5 \\ \bottomrule
   \end{tabular}
   \end{center}
   \flushright{\small $^*$ Training was unstable and experienced gradient explosion. \hspace{0.4cm}}
   \vspace{-1em}
\end{table}

Table~\ref{table:ablation} shows the results of an ablation study in which (a) we vary the number of message-passing iterations (left) and (b) disable various features of our architecture design (right).  As we can see from the left table, accuracy continues to improve as we increase the number of message passing iterations as high as 6; this gain indicates that our network learns to perform inference over long distances. The right table shows the impact of several of our design choices on the overall result. For example, if we do not use \emph{Contextual} edges (resp. \emph{Logical} edges), overall accuracy drops by $14.5\%$ (resp. $25.8\%$). These drops indicate that both kinds of predicates are crucial for achieving good accuracy. We also see that the attention layer for {\sc NPair}  makes a significant difference  for both library and user-defined types. Finally, Simple Aggregation is a variant of \lambdaNet{} that uses a simpler aggregation operation which replaces the attention-based weighed sum in Eq~\ref{eq:aggregation} with a simple average. As indicated by the last row of Table~\ref{table:ablation} (right), attention-based aggregation makes a substantial difference for user-defined types.

\subsection{Comparison with JSNice}
\revSilent{Since JSNice~\citep{JSNice} cannot properly handle class definitions and user-defined types, for a meaningful comparison, we compared both tools' performance on top-level functions randomly sampled from our test set. We filtered out functions whose parameters are not library types and manually ensured that all all the dependency definitions are also included. In this way, we constructed a small benchmark suite consisting of 41 functions. Among the 107 function parameter and return type annotations, \lambdaNet{} correctly predicted 77 of them, while JSNice only got 48 of them right. These results suggest that \lambdaNet{} outperforms JSNice, even when evaluated only on the places where JSNice is applicable.}

\section{Related Work}

% \heading{Type Inference using Statistical Methods.} Previous works like \textsc{NL2Type} (\cite{malik2019nl2type}), and \textsc{DeepTyper} (\cite{DeepTyper}) used variants of sequence models to predict type annotations for JavaScript programs. However, they do not take full advantage of the structure of the program. \textsc{DeepTyper} simply takes tokens as input, and \textsc{NL2Type} mostly uses information from documentation\jiayi{Mention NL2Type separately, as they are based on type information from document string}. In contrast, our model transforms the program to a \pgraph and performs structured learning over it to predict types. Another tool, \textsc{JSNice} (\cite{JSNice}), does perform structured prediction on a similar graph \jiayi{How similar is this?} using CRFs and MAP inference. Our work, however, takes advantage of deep learning using graph neural networks. Furthermore, our model is not limited to a fixed vocabulary of types.

\heading{Type Inference using Statistical Methods.} 
There are several previous works on predicting likely type annotations for dynamically typed languages: \cite{JSNice} and \cite{xu2016python} use structured inference models for Javascript and Python, but their approaches do not take advantage of deep learning and are limited to a very restricted prediction space. \cite{DeepTyper} and \cite{jangda2019predicting} model programs as sequences and AST trees and apply deep learning models (RRNs and Tree-RNNs) for TypeScript and Python programs. \cite{malik2019nl2type} make use of a different source of information and take documentation strings as part of their input. However, all these previous works are limited to predicting types from a fixed vocabulary.

% \heading{Graph Neural Networks for programs} We build upon the work of \cite{allamanis2017learning}. We specialize the model for the task of type prediction and introduce a new form of attention in the network, and a pointer-network like addition towards this goal. 

\heading{Graph Embedding of Programs.} \cite{allamanis2017learning} are the first to use GNNs to obtain deep embedding of programs, but they focus on predicting variable names and misuses for \cSharp{} and rely on static type information to construct the program graph. \cite{premiseSelectionEmbedding} use GNNs to encode mathematical formulas for premise selection in automated theorem proving. The way we encode types has some similarity to how they encode quantified formulas, but while their focus is on higher-order formulas, our problem requires encoding object types. \cite{GAT} are the first to use an attention mechanism in GNNs. While they use attention to compute node embeddings from messages, we use attention to compute certain messages from node embeddings.

\heading{Predicting from an Open Vocabulary.} Predicting unseen labels at test time poses a challenge for traditional machine learning methods. For computer vision applications, solutions might involve looking at object attributes \citep{farhadi2009} or label similarity \cite{wang_zero-shot_2018}; for natural language, similar techniques are applied to generalize across semantic properties of utterances \citep{dauphin2013zero}, entities \citep{eshel2017named}, or labels \citep{ren2016label}. Formally, most of these approaches compare an embedding of an input to some embedding of the label; what makes our approach a pointer network \citep{vinyals_pointer_2015} is that our type encodings are derived during the forward pass on the input, similar to unknown words for machine translation \citep{gulcehre_pointing_2016}.

\section{Conclusions}

We have presented \lambdaNet{}, a neural architecture for type inference that combines the strength of explicit program analysis with graph neural networks. \lambdaNet{} not only outperforms other state-of-the-art tools when predicting library types, but can also effectively predict user-defined types that have not been encountered during training. Our ablation studies demonstrate the usefulness of our proposed logical and contextual hyperedges. 

For future work, there are several potential improvements and extensions to our current system. One limitation of our current architecture is the simplified treatment of function types and generic types (i.e., collapsing them into their non-generic counterparts). Extending the prediction space to also include structured types would allow us to make full use of the rich type systems many modern languages such as TypeScript provide. Another important direction is to enforce hard constraints during inference such that the resulting type assignments are guaranteed to be consistent.

\section*{Acknowledgments}
We would like to thank DeepTyper authors, Vincent J. Hellendoorn, Christian Bird, Earl T. Barr, and Miltiadis Allamanis, for sharing their data set and helping us set up our experimental comparisons.
We also thank the ICLR reviewers for their insightful comments and constructive suggestions. Finally, we would also like to thank Shankara Pailoor, Yuepeng Wang, Jocelyn Chen, and other UToPiA group members for their kind support and useful feedback. This project was supported in part by NSF grant CCF-1762299.

\bibliography{lambdaNet.bbl}

\begin{thebibliography}{28}
\providecommand{\natexlab}[1]{#1}
\providecommand{\url}[1]{\texttt{#1}}
\expandafter\ifx\csname urlstyle\endcsname\relax
  \providecommand{\doi}[1]{doi: #1}\else
  \providecommand{\doi}{doi: \begingroup \urlstyle{rm}\Url}\fi

\bibitem[nd4()]{nd4j}
Deeplearning4j.
\newblock \url{https://github.com/eclipse/deeplearning4j}.
\newblock Accessed: 2019-09-24.

\bibitem[Allamanis et~al.(2017)Allamanis, Brockschmidt, and
  Khademi]{allamanis2017learning}
Miltiadis Allamanis, Marc Brockschmidt, and Mahmoud Khademi.
\newblock Learning to represent programs with graphs.
\newblock \emph{ICLR}, 2017.

\bibitem[Ancona \& Zucca(2004)Ancona and Zucca]{ancona2004principal}
Davide Ancona and Elena Zucca.
\newblock Principal typings for java-like languages.
\newblock In \emph{ACM SIGPLAN Notices}, volume~39, pp.\  306--317. ACM, 2004.

\bibitem[Bierman et~al.(2014)Bierman, Abadi, and
  Torgersen]{understanding-typescript}
Gavin Bierman, Mart{\'\i}n Abadi, and Mads Torgersen.
\newblock Understanding typescript.
\newblock In Richard Jones (ed.), \emph{ECOOP 2014 -- Object-Oriented
  Programming}, pp.\  257--281, Berlin, Heidelberg, 2014. Springer Berlin
  Heidelberg.
\newblock ISBN 978-3-662-44202-9.

\bibitem[Chung et~al.(2018)Chung, Li, Nardelli, and Vitek]{chung2018kafka}
Benjamin Chung, Paley Li, Francesco~Zappa Nardelli, and Jan Vitek.
\newblock Kafka: Gradual typing for objects.
\newblock In \emph{ECOOP 2018-2018 European Conference on Object-Oriented
  Programming}, 2018.

\bibitem[Dauphin et~al.(2013)Dauphin, Tur, Hakkani-Tur, and
  Heck]{dauphin2013zero}
Yann Dauphin, Gokhan Tur, Dilek~Z. Hakkani-Tur, and Larry~P. Heck.
\newblock Zero-shot learning for semantic utterance classification.
\newblock In \emph{ICLR}, 2013.

\bibitem[Eshel et~al.(2017)Eshel, Cohen, Radinsky, Markovitch, Yamada, and
  Levy]{eshel2017named}
Yotam Eshel, Noam Cohen, Kira Radinsky, Shaul Markovitch, Ikuya Yamada, and
  Omer Levy.
\newblock Named entity disambiguation for noisy text.
\newblock In \emph{CoNLL}, 2017.

\bibitem[Farhadi et~al.(2017)Farhadi, Endres, Hoiem, and Forsyth]{farhadi2009}
Ali Farhadi, Ian Endres, Derek Hoiem, and David Forsyth.
\newblock Describing objects by their attributes.
\newblock In \emph{CVPR}, 2017.

\bibitem[Gao et~al.(2017)Gao, Bird, and Barr]{type-or-not}
Zheng Gao, Christian Bird, and Earl~T. Barr.
\newblock To type or not to type: Quantifying detectable bugs in javascript.
\newblock In \emph{Proceedings of the 39th International Conference on Software
  Engineering}, ICSE '17, pp.\  758--769, Piscataway, NJ, USA, 2017. IEEE
  Press.
\newblock ISBN 978-1-5386-3868-2.
\newblock \doi{10.1109/ICSE.2017.75}.
\newblock URL \url{https://doi.org/10.1109/ICSE.2017.75}.

\bibitem[Gulcehre et~al.(2016)Gulcehre, Ahn, Nallapati, Zhou, and
  Bengio]{gulcehre_pointing_2016}
Caglar Gulcehre, Sungjin Ahn, Ramesh Nallapati, Bowen Zhou, and Yoshua Bengio.
\newblock Pointing the unknown words.
\newblock In \emph{Proceedings of the ACL}, 2016.

\bibitem[Hanenberg et~al.(2013)Hanenberg, Kleinschmager, Robbes, Tanter, and
  Stefik]{type-maintain}
Stefan Hanenberg, Sebastian Kleinschmager, Romain Robbes, {\'E}ric Tanter, and
  Andreas Stefik.
\newblock An empirical study on the impact of static typing on software
  maintainability.
\newblock \emph{Empirical Software Engineering}, 19:\penalty0 1335--1382, 2013.

\bibitem[Hellendoorn et~al.(2018)Hellendoorn, Bird, Barr, and
  Allamanis]{DeepTyper}
Vincent~J. Hellendoorn, Christian Bird, Earl~T. Barr, and Miltiadis Allamanis.
\newblock Deep learning type inference.
\newblock In \emph{Proceedings of the 2018 26th ACM Joint Meeting on European
  Software Engineering Conference and Symposium on the Foundations of Software
  Engineering}, ESEC/FSE 2018, pp.\  152--162, New York, NY, USA, 2018. ACM.
\newblock ISBN 978-1-4503-5573-5.
\newblock \doi{10.1145/3236024.3236051}.
\newblock URL \url{http://doi.acm.org/10.1145/3236024.3236051}.

\bibitem[Jangda \& Anand(2019)Jangda and Anand]{jangda2019predicting}
Abhinav Jangda and Gaurav Anand.
\newblock Predicting variable types in dynamically typed programming languages.
\newblock \emph{arXiv preprint arXiv:1901.05138}, 2019.

\bibitem[Kingma \& Ba(2014)Kingma and Ba]{Adam}
Diederik~P. Kingma and Jimmy Ba.
\newblock Adam: A method for stochastic optimization.
\newblock In \emph{ICLR}, 2014.

\bibitem[Li et~al.(2016)Li, Tarlow, Brockschmidt, and Zemel]{Gated-GNN}
Yujia Li, Daniel Tarlow, Marc Brockschmidt, and Richard~S. Zemel.
\newblock Gated graph sequence neural networks.
\newblock \emph{ICLR}, abs/1511.05493, 2016.

\bibitem[Malik et~al.(2019)Malik, Patra, and Pradel]{malik2019nl2type}
Rabee~Sohail Malik, Jibesh Patra, and Michael Pradel.
\newblock Nl2type: inferring javascript function types from natural language
  information.
\newblock In \emph{Proceedings of the 41st International Conference on Software
  Engineering}, pp.\  304--315. IEEE Press, 2019.

\bibitem[Mou et~al.(2016)Mou, Li, Zhang, Wang, and Jin]{mou2016convolutional}
Lili Mou, Ge~Li, Lu~Zhang, Tao Wang, and Zhi Jin.
\newblock Convolutional neural networks over tree structures for programming
  language processing.
\newblock In \emph{AAAI}, volume~2, pp.\ ~4, 2016.

\bibitem[Pierce \& Turner(2000)Pierce and Turner]{pierce2000local}
Benjamin~C Pierce and David~N Turner.
\newblock Local type inference.
\newblock \emph{ACM Transactions on Programming Languages and Systems
  (TOPLAS)}, 22\penalty0 (1):\penalty0 1--44, 2000.

\bibitem[Raychev et~al.(2015)Raychev, Vechev, and Krause]{JSNice}
Veselin Raychev, Martin Vechev, and Andreas Krause.
\newblock Predicting program properties from "big code".
\newblock In \emph{Proceedings of the 42Nd Annual ACM SIGPLAN-SIGACT Symposium
  on Principles of Programming Languages}, POPL '15, pp.\  111--124, New York,
  NY, USA, 2015. ACM.

\bibitem[Ren et~al.(2016)Ren, He, Qu, Voss, Ji, and Han]{ren2016label}
Xiang Ren, Wenqi He, Meng Qu, Clare~R Voss, Heng Ji, and Jiawei Han.
\newblock Label noise reduction in entity typing by heterogeneous partial-label
  embedding.
\newblock In \emph{Proceedings of the 22nd ACM SIGKDD International Conference
  on Knowledge Discovery and Data Mining}, pp.\  1825--1834. ACM, 2016.

\bibitem[Siek \& Taha(2007)Siek and Taha]{Siek2007GradualTF}
Jeremy~G. Siek and Walid Taha.
\newblock Gradual typing for objects.
\newblock In \emph{ECOOP}, 2007.

\bibitem[Traytel et~al.(2011)Traytel, Berghofer, and Nipkow]{hindley-milner}
Dmitriy Traytel, Stefan Berghofer, and Tobias Nipkow.
\newblock Extending hindley-milner type inference with coercive structural
  subtyping.
\newblock In \emph{Asian Symposium on Programming Languages and Systems}, pp.\
  89--104. Springer, 2011.

\bibitem[Veli{\v{c}}kovi{\'{c}} et~al.(2018)Veli{\v{c}}kovi{\'{c}}, Cucurull,
  Casanova, Romero, Li{\`{o}}, and Bengio]{GAT}
Petar Veli{\v{c}}kovi{\'{c}}, Guillem Cucurull, Arantxa Casanova, Adriana
  Romero, Pietro Li{\`{o}}, and Yoshua Bengio.
\newblock {Graph Attention Networks}.
\newblock \emph{International Conference on Learning Representations}, 2018.
\newblock accepted as poster.

\bibitem[Vinyals et~al.(2015)Vinyals, Fortunato, and
  Jaitly]{vinyals_pointer_2015}
Oriol Vinyals, Meire Fortunato, and Navdeep Jaitly.
\newblock Pointer networks.
\newblock In \emph{NeurIPS}, 2015.

\bibitem[Vitousek et~al.(2014)Vitousek, Kent, Siek, and
  Baker]{vitousek2014design}
Michael~M Vitousek, Andrew~M Kent, Jeremy~G Siek, and Jim Baker.
\newblock Design and evaluation of gradual typing for python.
\newblock In \emph{ACM SIGPLAN Notices}, volume~50, pp.\  45--56. ACM, 2014.

\bibitem[Wang et~al.(2017)Wang, Tang, Wang, and
  Deng]{premiseSelectionEmbedding}
Mingzhe Wang, Yihe Tang, Jian Wang, and Jia Deng.
\newblock Premise selection for theorem proving by deep graph embedding.
\newblock In \emph{Advances in Neural Information Processing Systems}, pp.\
  2786--2796, 2017.

\bibitem[Wang et~al.(2018)Wang, Ye, and Gupta]{wang_zero-shot_2018}
Xiaolong Wang, Yufei Ye, and Abhinav Gupta.
\newblock Zero-shot recognition via semantic embeddings and knowledge graphs.
\newblock In \emph{CVPR}, 2018.

\bibitem[Xu et~al.(2016)Xu, Zhang, Chen, Pei, and Xu]{xu2016python}
Zhaogui Xu, Xiangyu Zhang, Lin Chen, Kexin Pei, and Baowen Xu.
\newblock Python probabilistic type inference with natural language support.
\newblock In \emph{Proceedings of the 2016 24th ACM SIGSOFT International
  Symposium on Foundations of Software Engineering}, pp.\  607--618. ACM, 2016.

\end{thebibliography}
\bibliographystyle{iclr2020_conference}

% \appendixc
% \section{Appendix}

\end{document}